\documentclass[
reprint,
superscriptaddress,
 amsmath,amssymb,
 aps,
prstab,
floatfix,
]{revtex4-1}

\usepackage{graphicx}
\usepackage{dcolumn}
\usepackage{bm}
\usepackage[english]{babel}
\usepackage[utf8]{inputenc}
\usepackage{siunitx}
\usepackage[dvipsnames]{xcolor}
\usepackage{hyperref}
\hypersetup{
    colorlinks=true,
    linkcolor=blue,
    filecolor=magenta,      
    urlcolor=cyan,
}
\makeatletter
\renewcommand{\fnum@figure}{\normalfont FIG. \thefigure}
\renewcommand*{\@caption@fignum@sep}{\normalfont . }
\makeatother

\begin{document}

\preprint{APS/123-QED}

\title{Characterizing ultra-low emittance electron beams using structured light fields}
\author{Andreas Seidel}
\email{seidel.andreas@uni-jena.de}
\affiliation{Friedrich-Schiller-Universit{\"a}t, F{\"u}rstengraben 1, 07743 Jena, Germany}
\affiliation{Helmholtz-Institut Jena, Fröbelstieg 3, 07743 Jena, Germany}
\author{Jens Osterhoff}
\affiliation{Deutsches Elektronen-Synchrotron DESY, Notkestraße 85, 22607 Hamburg, Germany}
\author{Matt Zepf}
\affiliation{Friedrich-Schiller-Universit{\"a}t, F{\"u}rstengraben 1, 07743 Jena, Germany}
\affiliation{Helmholtz-Institut Jena, Fröbelstieg 3, 07743 Jena, Germany}
\date{\today}

\begin{abstract}
Novel schemes for generating ultra-low emittance electron beams have been developed in the last years and promise compact particle sources with excellent beam quality suitable for future high-energy physics experiments and free-electron lasers. Current methods for the characterization of low emittance electron beams such as pepperpot measurements or beam focus scanning are limited in their capability to resolve emittances in the sub $0.1$ mm mrad regime. Here we propose a novel, highly sensitive method for the single shot characterization of the beam waist and emittance using interfering laser beams. In this scheme, two laser pulses are focused under an angle creating a grating-like interference pattern. When the electron beam interacts with the structured laser field, the phase space of the electron beam becomes modulated by the laser ponderomotive force and results in a modulated beam profile after further electron beam phase advance, which allows for the characterization of ultra-low emittance beams. 2D PIC simulations show the effectiveness of the technique for normalized emittances in the range of $\epsilon_n=[0.01,1]$ mm mrad. 
\end{abstract}
                      
\maketitle


\section{Introduction}

The emittance of an electron beam is one of its most important quality measures for many applications. The emittance describes the volume occupied by the electron beam in 6-dimensional phase-space \cite{Buon1992a}. The transverse emittance of the beam determines the smallest spot size that can be achieved for a given focusing arrangement and hence influences the luminosity in HEP experiments. Another area of importance is in determining the gain of free-electron lasers through the Pierce-parameter \cite{Barletta2010}.

In linear accelerators the injection of low emittance electron bunches into the accelerating structure is of prime importance. Significant improvements in radio-frequency (RF) guns with laser photo-cathodes allow a small initial source size and limit the emittance growth in the early stages using strong accelerating fields of several 10s of MV/m. Using this approach GeV level linear accelerators (linacs) currently achieve a normalised transverse emittance $\epsilon_n=\beta_e \gamma_e \epsilon$ ($\beta_e=v_e/c$, $\gamma_e$ - electron gamma factor and $\epsilon$ - geometric beam emittance) of $0.002$ mm mrad \cite{Ji2019Emittance} for a bunch with less than 10 electrons, $0.10$ mm mrad \cite{Marx2018Emittance} for a bunch with a charge of $2$ pC and  $0.89$ mm mrad \cite{Rimjaem2010} for a bunch with a charge of 1 nC, with continuous strides being made to achieve even lower emittances.

Laser plasma wakefield accelerators (LWFA)\cite{Esarey2009a,Downer2018} feature very large accelerating fields reaching 100 GV/m or more. In principle these large fields result in the electron beam becoming highly relativistic over a propagation length of a fraction of a mm and consequently provide a promising route to achieving extremely low emittances. As in the case of RF accelerators, the injection volume is a key determinant of the final emittance. This is especially true for low charge beams, where emittance growth due to space charge is negligible. To date most laser plasma wakefield experiments with a dedicated injection procedure have used either ionisation-\cite{Pak2010_ionisationInjection,Guffey2010IonizationInjection}, downramp-\cite{PhysRevLett.94.115003} or colliding pulse injection \cite{Faure2006_collidingPulse} and  transverse emittances of similar magnitude to RF linear accelerators have been attained ($\epsilon_n<0.2$ mm mrad  \cite{Weingartner2012a,PhysRevLett.109.064802}). 

Further improvements can be realized if the  phase space injection volume and field perturbations are minimised. A promising approach here is the plasma photocathode, which combines the well-defined volume achievable by ionisation injection with a separate injection laser with a particle-beam-driven wakefield \cite{Hidding2012}. This approach allows the laser intensity used for ionisation injection to be minimised resulting in a low initial momentum spread.  Additionally, emittance growth is reduced by eliminating the powerful laser driving the wakefield structure and thus the unavoidable field fluctuations introduced by an oscillating driver field. Simulations have predicted that an emittance of the order of 10 $\mu$m mrad is feasible \cite{LiTransverseColliding2013}.

The potential performance gains by such a marked reduction in emittance are highly desirable and experiments are underway to explore the development of lower emittance beams. Current measurement techniques which are being used for mono and poly-chromatic electron beams such as beam focus scanning have shown to be capable of measuring emittances down to $\epsilon_n \approx 0.2$ mm mrad \cite{Weingartner2012a,PhysRevLett.109.064802}. A further improvement of the detection limit using beam focus scanning techniques is conceivable. However, their practical realization is quickly becoming challenging for smaller emittances and large setups are required. A novel knife-edge based measurement characterization is capable of measuring electron beam emittances down to $0.002$ mm mrad \cite{Ji2019Emittance}, but was limited to mono energetic and low energy ($<$1MeV) electron beams. Hence new approaches are needed to characterize the next generation of electron beams. 

\begin{figure*}[t]
	\includegraphics[width=0.9\linewidth]{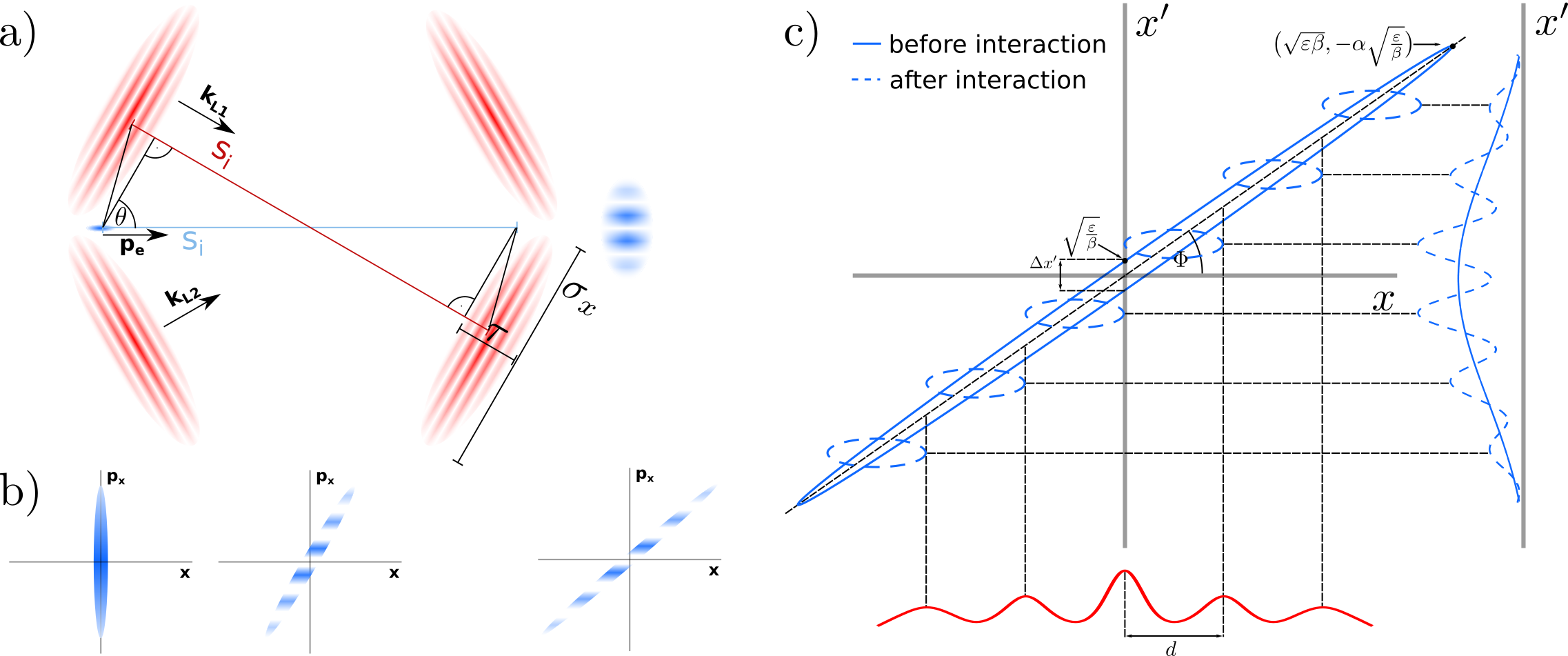}
	\caption{(a) Schematic of the measurement setup showing two laser beams crossing under an angle $2\cdot(90-\theta)$ to form an interference pattern in the interaction region(Fig. \ref{fig:focus}). After the laser-electron-interaction the imparted momentum modulation evolves into a transverse density modulation that can be observed on a downstream beam monitor screen and allows the e-beam waist size to be determined. Inclusion of a magnet behind interaction allows the emittance of monochromatic beam slices to be measured even for beams with a large projected energy spread. (b) Schematic of the beam momentum distribution at the waist, the interaction point and after free space propagation (from left to right). (c) Transverse electron phase space before (solid) and after laser-electron interaction (dashed). The red line indicates the intensity profile of the two interfering laser beams at the corresponding transverse position x at the interaction point. The ponderomotive force shifts the transverse momentum of the electrons. For an optimum laser intensity the momentum transfer is matched to the gradient of the phase-space and shifts particles from regions of high laser intensity to regions of low intensity highlighted by the dashed ellipses.}
	\label{fig:Setup}
\end{figure*}

In the following we describe a method that spatially modulates the transverse momentum distribution of the electron beam on $\mu$m spatial scale using a laser beam and allows the size of the electron beam waist to be determined from the observed modulation on a simple beam monitor screen downstream. The sensitivity can be adjusted via the spatial frequency of the laser interference pattern, allowing e-beam waists from 10s of nm to $\mu$m-scale to be determined. 
The emittance range that can be covered by this technique is also determined by the divergence of the e-beam and is compatible with a large emittance parameter range from the mm mrad  of current linacs to the $\mu$m mrad anticipated for novel injection schemes. Measurements of small spatial scales with laser interference structures is a viable approach to assess the quality of electron beams and has been demonstrated in beam-size monitors, where in contrast to wire scanners the fine wire is replaced by a 'laser-wire' formed by interfering lasers in the so called Shintake-monitor \cite{Yan2012a}. It can resolve nm-spatial scale electron beam foci using a scanning technique. While this method requires a measurement of the electron bunches at focus, which can be an immense effort for poly-chromatic electron beams, our proposal is compatible with broadband diverging beams and also allows a determination of the slice emittance.

\section{Theoretical Model}

The proposed measurement scheme is shown schematically in Figure \ref{fig:Setup}. Two laser beams with wavelength $\lambda$ cross under an angle $2\cdot(90-\theta)$ to form an interference pattern with  a periodicity $d = \lambda/(2 \cos{\theta})$ at a distance $z$ from the electron beam waist (Fig. \ref{fig:focus}). In the absence of the laser beam the unperturbed electron beam profile is visible on a measurement screen at a convenient location downstream of the interaction region. When the laser is switched on, the transverse momenta of the electrons are modulated by the field of the laser with the periodicity of the interference pattern allowing a modulated electron beam pattern to be observed on the screen.

The working principle is shown in greater detail in Fig. \ref{fig:Setup}(c). The electron beam transverse dimension $x$ and divergence angle $x'$ can be described as an ellipse \cite{Floettmann2003} parameterized by TWISS parameters $ \alpha, \beta, \gamma$ so that $\epsilon=\gamma x^2 + 2\alpha x x' + \beta x'^2$. In this description the beam width $\sigma_x$  and angular spread $\sigma_{x'}$ are given by $ \sigma_x(z) = \sqrt{\beta(z)\epsilon}$ and $\sigma_{x'}(z)=\sqrt{\gamma(z)\epsilon}$ as a function of the distance $z$ from the beam waist respectively and $\tan \Phi =dx'/dx \propto 1/z$. Note that while the ellipse has an overall width of $\sigma_{x'}$ the slice width at a given transverse position $x$ can be significantly less $\Delta x'= 2\sqrt{\frac{\epsilon}{\beta(z)}}$ and decreases with the propagation distance z from the waist. This implies that a visible modulation of the electron beam can be achieved by imparting a transverse kick to the electrons of  $\sim\Delta x'$, in which case the electron beam will exhibit areas of increased phase space density at certain values $x'$ and reduced at others.  As can be seen from Fig. \ref{fig:Setup}(c), the separation of adjacent intensity peaks $d$ must be be large enough to prevent the two adjacent regions with slice width $\Delta x'=2 \sqrt{\epsilon/\beta(z)}$ from overlapping in $x'$-space. At large values of $\alpha$ (i.e. far from the waist)  the lower limit for the laser interference pattern period $d$ is therefore:
\begin{figure}[t]
	\centering
	\includegraphics[width=0.7\linewidth,trim={0.4cm 0.5cm 0.2cm 0.3cm},clip]{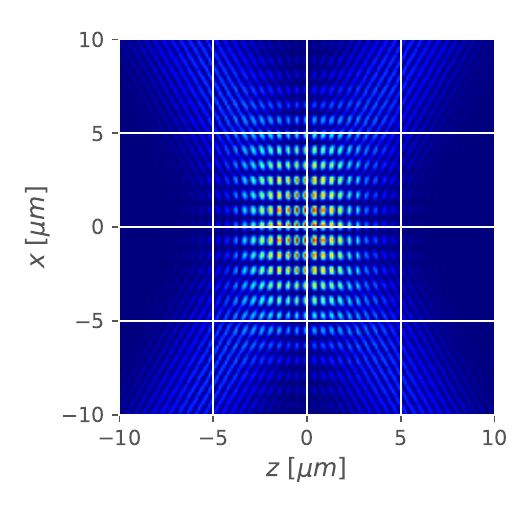}
	\caption{Intensity pattern of the interference of the colliding laser pulses.}
	\label{fig:focus}
\end{figure} 
\begin{align}
    d> 2\frac{\sqrt{\epsilon \beta(z)}}{\alpha(z)}\approx 2 \frac{\epsilon}{\sigma_{x'}}=2 \frac{\epsilon}{\sigma_{x'}}= 2\sigma_x(0)=const
    \label{eq:interference_d_min}
\end{align}
The condition for the minimum periodicity is therefore connected to the beam waist size and this condition is independent of the distance z from the electron beam waist where the measurement is taken. As we shall see in the following, the waist size can be determined from the modulation depth without the need for a direct measurement at the location of the beam waist itself. Combined with a simple measurement of the  unperturbed beam divergence we can therefore use this method to accurately determine the emittance of the electron beam.

The laser imparts momentum to the electrons via the ponderomotive force $F_p \propto \nabla E^2$ of the laser which tends to push electrons from regions of high intensity to regions of low intensity. As can be seen from Fig. \ref{fig:Setup}(c) the optimal modulation depth is achieved when electrons are shifted in transverse momentum by an amount that matches the gradient of the ellipse in phase space. In momentum terms this requirement can be expressed by 
\begin{align}\label{eq:interference_pattern_impulse}
        \frac{\Delta p_x}{\Delta x}=\frac{dp_x}{dx} = p_z \tan{\Phi} \propto \frac{p_z}{z}
\end{align}
Assuming that the ponderomotive force is the dominant force in changing the particle momentum, we can derive the requirement 
\begin{align*}
        \frac{dp_x}{dtdx} \overset{!}{=} \frac{F_{pond}}{dx}
\end{align*}
For a given gradient of the transverse momentum across the electron beam, an optimal laser intensity leads to the desired modulation of transverse momentum and, hence, maximum modulation visibility. The change in transverse momentum can be expressed in terms of the interaction time $t_{int}$ and laser intensity $I$ as 

\begin{align}\label{eq:interference_pattern_intensity}
        \frac{dp_x}{dx} = \frac{\langle F_{pond}\rangle t_{int}(\gamma_e)}{dx} \propto \frac{d\langle I \rangle }{\omega^2\hspace{1mm}dx}\frac{t_{int}(\gamma_e)}{dx} ,
\end{align}
where $\langle \rangle$ is a time average over the duration of the interaction and $\omega$ the laser frequency. 
Since the intensity of the interference pattern drops from its maximum to zero over a distance of $d/2$ and using Eq.(\ref{eq:interference_pattern_intensity}) we obtain:
\begin{align*}
        \frac{\Delta p_x}{\Delta x} \propto \langle I \rangle_0 \cdot t_{int}(\gamma_e),
\end{align*}
 where $\langle I \rangle_0$ is the time averaged intensity required to achieve the optimum beam modulation. Inserting in to Eq.(\ref{eq:interference_pattern_impulse}) we find the average interaction intensity to achieve the optimum modulation depth with:
\begin{align*}
        \langle I \rangle_0 \propto \frac{p_z}{z \cdot t_{int}(\gamma_e)} 
\end{align*}
   
 As one would  expect, the laser intensity increases for higher energy electron beams and decreases with increasing distance from the beam waist. The latter scaling can be simply understood, as the increased beam size with increasing $z$ reduces the gradient of transverse momentum. This  corresponds to a  shallower gradient of the emittance ellipse and a reduced  momentum modulation required to be imparted by  the laser. Note that the required laser energy also depends on the interaction length (ignoring beam size effects) $s_i=c\tau/(1-\sin(\theta))$, favouring shallow crossing angles if a sufficiently small $d$ is maintained. The required intensity can thus be kept sub relativistic, which diminishes the demands on the laser system.

Assuming an electron bunch with Gaussian transverse beam waist $\sigma_{x0}$ and corresponding momentum $\sigma_{px0}$ we obtain 
\begin{align}
    n(x,p_x) = n_{e0} \cdot exp \left(-\frac{p_x^2}{2\sigma_{px0}^2}-\frac{x^2}{2\sigma_{x0}^2}\right )
\end{align}
for the density distribution at waist position. With Eq.(\ref{eq:interference_pattern_impulse}) and (\ref{eq:interference_pattern_intensity}) we obtain 
\begin{align}
\begin{split}
      n(p_x) = n_{e0} \int_{-\infty}^\infty & exp \bigg( -\frac{(p_x)^2}{2\sigma_{px0}^2} \\
      & -\frac{((p_x-\Delta p_x)/(p_z\tan{\Phi})-x)^2}{2\sigma_{x0}^2}\bigg) dx,
\end{split}
\label{eq:electron_momentumspace}
\end{align}
for the momentum space after the laser-electron interaction, where $\Delta p_x=\langle F_{pond} \rangle \cdot t_{int}(\gamma_e)$. The first term in the exponent represents the transverse divergence of the electron bunch, which is not affected by the interaction with the laser interference pattern. The second term describes the changes in the transverse momentum space caused by the ponderomotive force and indicates an increase in the peak height of the created modulations with decreasing values of $\sigma_{x0}$.

\section{Simulation Results}
2D simulations to test the scheme were conducted using the  code EPOCH \cite{Arber2015}. The simulations were performed in a moving box with the size of $z=320$ $\mu$m and $x=80$ $\mu$m with 20 cells/micron resolution in every dimension. The  electron bunch, sampled by $8\cdot 10^5$ macro particles, had a longitudinal size of $\sigma_z=2$ $\mu$m and variable transverse size $\sigma_x$ and propagated along the z-axis. The electron bunch charge was chosen at 150 fC to ensure an interaction between laser and electron bunch without further charge related effects. The interaction laser was modelled as two pulses propagating linearly polarized in y-direction and with a duration $\tau_{FWHM}=17$ fs, spot size at interaction $\sigma_{FWHM}=30$ $\mu$m at a wavelength of  $\lambda=0.8$ $\mu$m. The laser intensity distribution  produced by the two  interfering pulses is shown in Fig.  \ref{fig:focus} with the lasers propagating from the upper left side and the lower left side. The angle between z-axis and lasers is $90-\theta=30^\circ$.

\begin{figure}[t]
	\includegraphics[width=1\linewidth,trim={0.7cm 0.5cm 0.6cm 0.3cm},clip]{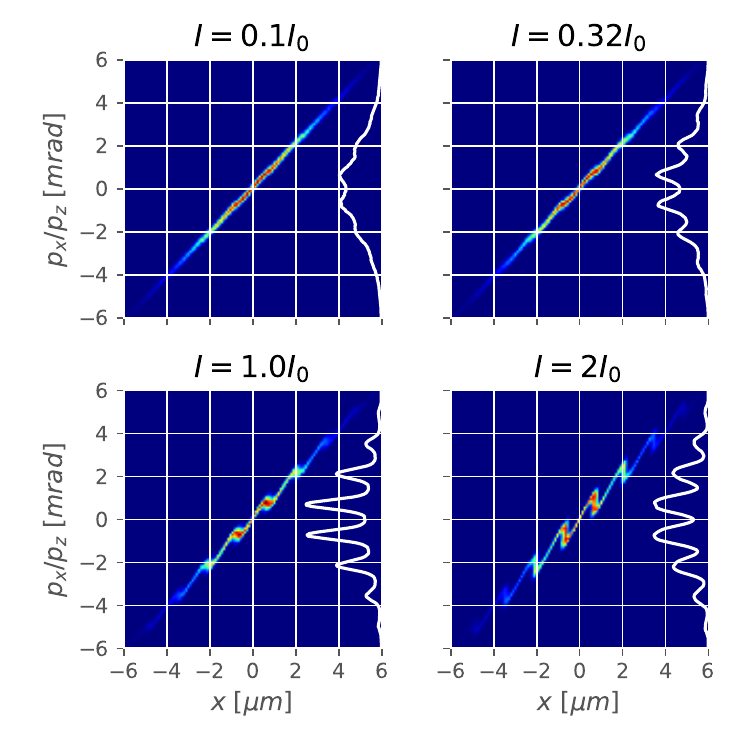}
	\caption{Momentum space of electron beam ($\gamma_e = 100$, $ \sigma_x=100$ nm, $\sigma_{x'}=2$ mrad, $\sigma_z=2$ $\mu$m) after interaction with different laser intensities. As can be seen the modulation optimizes for a specific intensity and reduces again for higher values. White curve: transverse momentum space integrated along x (measured signal on detector).}
	\label{fig:emittance_vs_intensity}
\end{figure} 

Fig. \ref{fig:emittance_vs_intensity} shows the electron beam modulations for different laser intensities. The initial electron bunch parameters are $\gamma_e = 100$, $ \sigma_x=100$ nm, $\sigma_{x'}=2$ mrad, $\sigma_z=2$ $\mu$m and $I_0 = 3.5\cdot 10^{16}$ W/cm$^2$. The  ponderomotive force imparts transverse momentum to the electrons and concentrates the electrons at certain propagation angles corresponding to the nodes of the interference pattern, resulting in an intensity modulation on the diagnostic.
Although these laser intensities do not cause instantaneous changes in the position of the electrons, the electrons drift due to their transverse momentum during interaction in the interference grating. This finally leads to a blurring of the modulation signal on the detector. Our simulations have shown for an electron drift during the interaction of less than $10\%$ of the laser interference pattern period d, there is no measurable difference in the modulation. 

\begin{figure}[t]
	\includegraphics[width=1\linewidth,trim={0.5cm 0.4cm 0.3cm 0.3cm},clip]{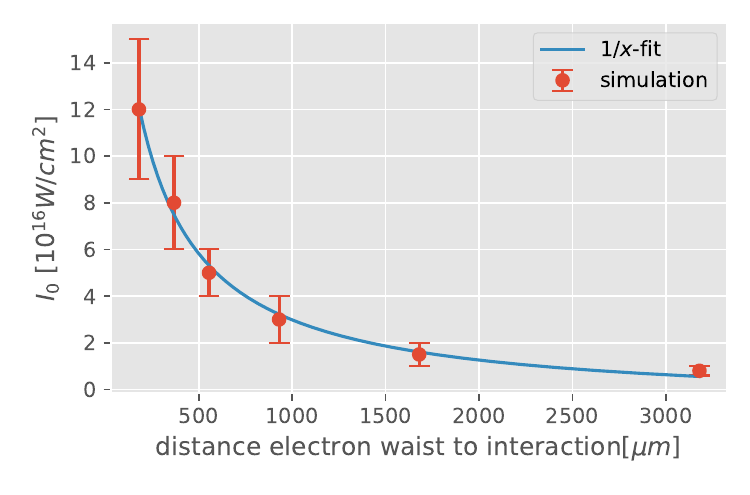}
	\caption{Optimum peak Intensity of a single laser used, to get strongest modulation signal in momentum space of the electron beam($\gamma_e = 100$,  $\sigma_x=150$ nm, $\sigma_{x'}=2$ mrad, $\sigma_z=2$ $\mu$m) for various distances between electron beam being at waist and laser-electron interaction. The vertical bars indicate the intensity range over which the change in maximum modulation depth is less than $5\%$.}
	\label{fig:intensity_vs_time}
\end{figure}

The modulation is strongest for the matched intensity $I_0$ and decreases above and below this. For intensities that are too low, the perturbation is much smaller than the local slice angular spread  $\Delta x'$ (the ellipse-width in $x'$-direction) leading to a negligible effect. At large intensities the electrons are displaced by much more than the local ellipse-width and the effect is 'smeared out'. 
For a given set of experimental parameters, the peak laser intensity at which the optimal modulation depth occurs is shown in Figure \ref{fig:intensity_vs_time}.  With $P_{Laser} \approx I_0/2\cdot\pi \sigma^2$ the required laser power for this particular electron bunch and interaction point $1600$ $\mu$m behind the electron waist matched to a laser spot size $\sigma_{FWHM}=30$ $\mu$m  is $300$ GW. 

One might assume that this method places increasingly onerous requirements on the laser to make a measurement for beams with very high electron beam energy. Since the width of the ellipse and therefore the optimal intensity depends on the distance z from the beam waist as $1/z$, allowing the optimal intensity to be controlled by the interaction geometry. Note that for limited available laser power there is no intrinsic requirement for the laser spot to be larger than the electron beam. In principle smaller spots can be used with scanning measurements to determine the emittance of the beam, therefore enabling measurements with lasers that are easily co-located with an electron beam. Note that any relative pointing jitter between laser interference pattern and electron bunch leads to the same results as the proposed scanning measurement. However, it must be ensured that the laser interference pattern remains in place until the whole electron bunch in longitudinal direction has interacted with it.

\begin{figure}[b]
	\includegraphics[width=1\linewidth,trim={0.5cm 0.4cm 0.2cm 0.5cm},clip]{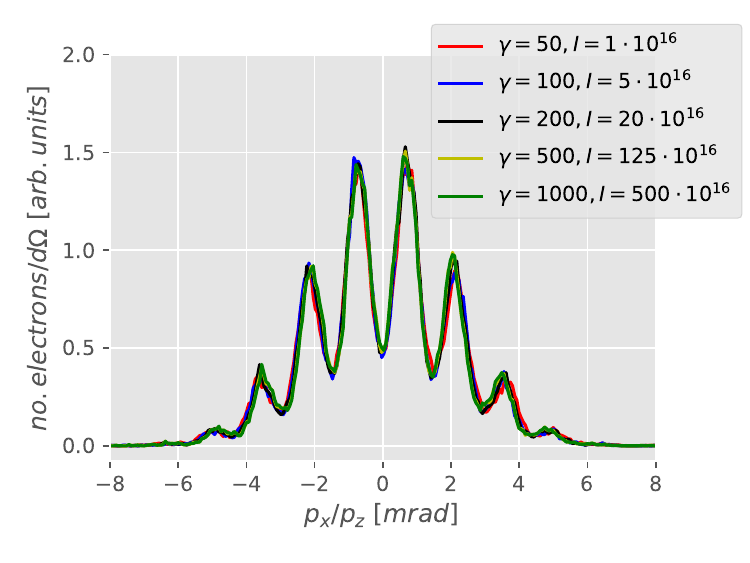}
	\caption{Transverse momentum space of electron beams with different $\gamma_e$ ($\sigma_{x'}=2$ mrad, $\sigma_z=2$ $\mu$m, $\sigma_x=$ 150nm) after interaction. Laser intensity is optimized for maximum modulation depth.}
	\label{fig:emittance_vs_gamma_e}
\end{figure}

As is clear from the discussion in the previous section and can be seen in Fig. \ref{fig:emittance_vs_gamma_e} higher energy electron beams require a higher laser intensity for optimal modulation for otherwise identical parameters. The increase is quadratic in the Lorentz-factor $\gamma_e$ due to the relativistic contraction of the interaction length and the higher transverse momentum required to achieve the same ratio of $p_x/p_z$ on the beam electrons. It is important to note, that at the optimum laser intensity the maximum modulation depth is independent of the electron bunch $\gamma_e$.
\begin{figure}[b]
	\includegraphics[width=1\linewidth,trim={0.6cm 0.5cm 0.5cm 0.3cm},clip]{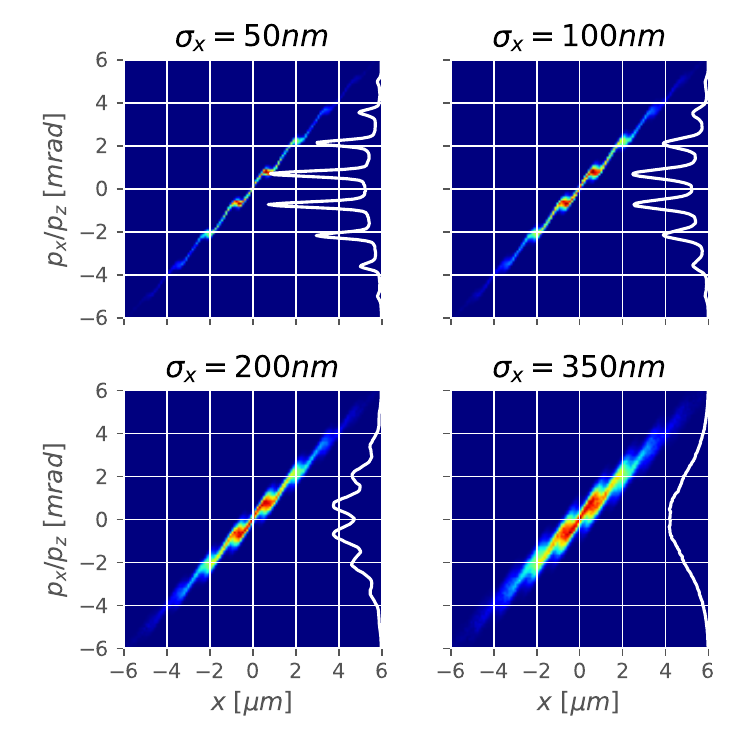}
	\caption{Momentum space of electron beam($\gamma_e = 100$, $\sigma_{x'}=2$ mrad, $\sigma_z=2$ $\mu$m) after interaction with laser ($I_0=3.5\cdot 10^{16}$ W/cm$^2$) for different electron source sizes $\sigma_x$. White curve: transverse momentum space integrated along x (measured signal on detector).}
	\label{fig:emittance_vs_sourcesize}
\end{figure}

From the discussion above we can now formulate a measurement approach for calculating the desired e-beam characteristics. As is clear from Figure \ref{fig:emittance_vs_sourcesize}
 the change in the beam intensity modulation depends only on source size for fixed laser intensity with the beam modulation visible as long as the criterion from Eq. \ref{eq:interference_d_min} is met. The proposed measurement strategy is therefore as follows. First a location at some distance z from the e-beam waist is chosen depending on the beam and available laser parameters. Varying the intensity of the laser allows the optimal laser intensity to be set by optimizing the modulation depth, thereby eliminating any systematic effects such as small offsets in achieved actual laser intensity from nominal or distance z from the electron beam waist. It should be noted here that deviations from the optimal laser intensity of $<10\%$ did not lead to measurable changes in the simulations peak height (see Fig. \ref{fig:intensity_vs_time}). At the optimum intensity and for a known interference period $d$ the modulation depth measured in a single shot allows the beam waist size to be directly inferred from the modulated beam profile by comparing with the pic-simulation/analytical solution which leads to the normalized emittance with $\epsilon_n = \beta_e\gamma_e \sigma_{px0}\sigma_{x0}$. Our analysis has shown this comparison also applies for a laser profile with minor deviations to the theoretically assumed Gauss profile. We compared versions of the analytical solution in which noise of up to $20\%$ was added to the ponderomotive force to the version without and observed no measurable difference in modulation depth ($\sigma_x>0.05\cdot d$).

\begin{figure}[t]
	\includegraphics[width=1\linewidth,trim={0.3cm 0.5cm 0.5cm 0.2cm},clip]{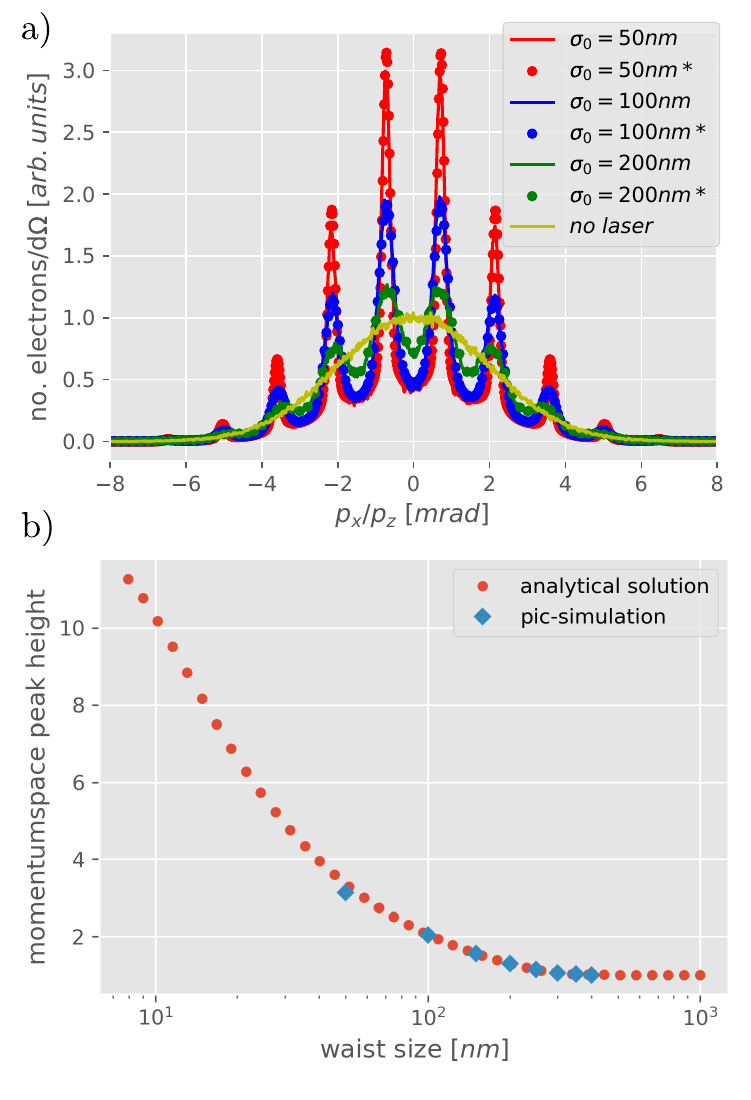}
	\caption{(a) Transverse momentum space of electron beams with different $\sigma_x$ ($\sigma_{x'}=2$ mrad, $\sigma_z=2$ $\mu$m, $\gamma_e=100$ and $d=0.8$ $\mu$m) after interaction. Solid line indicates the PIC simulation results and dashed the solution of Eq. \ref{eq:electron_momentumspace} assuming a sinusoidal electrical laser field. (b) Normalized peak height of modulation in transverse momentum space after laser-electron interaction calculated with Eq. \ref{eq:electron_momentumspace}. No interaction results in a height of 1.}
	\label{fig:waistsize_vs_modulation}
\end{figure}

 Figure \ref{fig:waistsize_vs_modulation} compares the calculation of the modulation from the analytical considerations above to the PIC simulations. Clearly the source size can be sensitively determined over a large range with a single measurement configuration. The measurement sensitivity can be further increased by reducing the interference period $d$. Note that for the smallest waist-size ($50$ nm) in the simulation, which corresponds to an emittance of $10$ $\mu$m mrad, an electron detection system with a resolution better than $40$ $\mu$rad would be required. 
 
This method is capable of characterising very small waist size beams with unprecedented emittance accurately. The sensitivity of the emittance measurement increases for lower divergence electron beams. Electron beam waists down to 10s of nm can be resolved using this technique corresponding to normalised emittance of the order of $\mu$m mrad. Using current laser technology it is possible to characterise GeV e-beams, thereby providing a precise technique to characterise ultra-low emittance electron beams under development.
In principle this technique can also accommodate beams with significant energy spread, such as those from wakefield accelerators by combining the beam modulation in one plane with a dipole magnet dispersing the electron beam in the other plane. Simulations for a poly-chromatic electron bunch with an energy spread of $< 5\%$  resulted in no measurable changes in the peak height, which agrees with the sensitivity of the modulation depth to deviations from the optimal intensity. We note that while this technique becomes more demanding in terms of laser intensity at higher electron beam $\gamma_e$ it can still be applied and required laser beam energy can be reduced by using sufficiently small spots combined with scanning measurements across the beam spatial dimensions.
\section{Conclusion}
We have described a novel scheme for characterising the properties of an electron beam. To our knowledge, the method is unique in allowing the measurement of extremely small broadband electron beam source sizes and emittances in the $\mu$m mrad regime predicted to be accessible using advanced accelerator techniques. Our simulations have  shown that emittances as small as  $\epsilon_n=0.01$ mm mrad can be well resolved. However, this value does not represent an obvious lower limit and can be further reduced by adjusting the parameters appropriately. We note that the method is suited for mono energetic bunches (from RF linacs) and  LWFA bunches with broader bandwidth. The high temporal resolution of the method has the potential to allow slice emittance to be determined for different parts of the beam along the propagation axis, given a monotonic time-energy correlation in longitudinal phase space and a laser pulse duration which is shorter than the electron bunch length. Cross calibration  with well-established emittance measurement methods is possible using typical Ti:Sapphire laser systems and electron beams with a transverse beam waist $\sigma_{x0} < 2$ $\mu$m.

\section{Acknowledgements}
This research was funded by the Federal Ministry
of Education and Research of Germany in the Verbundforschungsframework (project number 05K16SJB).

\bibliography{references}

\end{document}